\documentclass [11pt]{article}

\usepackage {graphicx}
\usepackage {longtable}
\begin{document}

\begin{center}
\bf{\LARGE Relativistic Mean-Field Theory Equation of State of
Neutron Star Matter and a Maxwellian Phase Transition to Strange
Quark Matter }\footnote{\textit{Astrophysics, vol.52, No.1, pp.
132-150, 2009. Translated from Astrofizika, 52, No.1, pp. 147-164,
2009. }}

\vskip10pt{\bf{G. B. Alaverdyan }
\vskip5pt{\textit {\small
Yerevan State University, A.Manoogyan str. 1, 0025 Yerevan,
Armenia
\\ E-mail: galaverdyan@ysu.am}}}
\end{center}

\begin{quote}
{\small{ \textbf{Abstract.}\textit{The equation of state of
neutron star matter is examined in terms of the relativistic
mean-field theory, including a scalar-isovector   $\delta$-meson
effective field. The constants of the theory are determined
numerically so that the empirically known characteristics of
symmetric nuclear matter are reproduced at the saturation density.
The thermodynamic characteristics of both asymmetric nucleonic
matter and $\beta$-equilibrium hadron-electron $npe$-plasmas are
studied. Assuming that the transition to strange quark matter is
an ordinary first-order phase transition described by Maxwell's
rule, a detailed study is made of the variations in the parameters
of the phase transition owing to the presence of a $\delta$-meson
field. The quark phase is described using an improved version of
the bag model, in which interactions between quarks are accounted
for in a one-gluon exchange approximation. The characteristics of
the phase transition are determined for various values of the bag
parameter within the range $B\in[60,120]$ $MeV/fm^{3}$ and it is
shown that including a   $\delta$-meson field leads to a reduction
in the phase transition pressure $P_{0}$ and in the concentrations
$n_{N}$ and $n_{Q}$ at the phase transition point.}

 }}
\end{quote}

\bigskip

\section{Introduction}
In addition to their independent, fundamental significance,
studies of the structural characteristics and composition of the
constituents of matter at extremely high densities and
temperatures play an extremely important role in clarifying the
physical nature of the internal structure and integral parameters
of neutron stars. A quantum field approach in the framework of
quantum hadrodynamics (QHD) provides a fairly adequate description
of the properties of nuclear matter and of finite nuclei, treating
them as a system of strongly interacting baryons and mesons. One
theory of this type that has effective applications, is the
relativistic mean-field theory \cite{Wal74,SWNV86,SW97}. This
theory yields satisfactory descriptions of the structure of finite
nuclei \cite{TW99}, the equation of state of nuclear matter
\cite{MS95}, and the features of heavy ion scattering
\cite{KoLi96}. The parameters of the mean-field model
characterizing the interaction of a nucleon with $\sigma$,
$\omega$, and $\rho$ mesons can be self consistently determined
starting with empirical data on symmetric nuclear matter
 near the saturation density. This, in turn, leads to the
possibility of obtaining equations of state for superdense,
isospin-asymmetric nuclear matter. In these studies it has been
assumed that the masses of the scalar-isoscalar ($\sigma$),
vector-isoscalar ($\omega$), and vector-isovector ($\rho$) mesons
and their coupling constants are independent of the density and of
the values of the fields. In addition, the scalar-isovector
$\delta$-meson ($a_{0}(980)$) is not included among the exchange
mesons.

Relativistic mean-field theory models have been constructed
\cite{Vosk05,Ryu05} assuming that the nucleon and exchange meson
masses in nuclear media obey the Brown-Rho scaling law
\cite{BR91}. The results showed that including the density
dependence of the mass leads to a more stiff equation of state for
the matter. A scalar-isovector $\delta$ -meson has been added to
the scheme and a study of its role for asymmetric nuclear matter
at low densities has been made in Refs.
\cite{Kub97,Liu02,GreCol03}. This approach has been used
\cite{Gre03,Gai04,Toro06} to study scattering processes for
neutron-reach heavy ions at medium energies and the feasibility of
forming a hadron-quark mixed phase during the collision process.

This paper is a study of the equation of state for neutron star
matter in terms of the relativistic mean-field theory and an
examination of the variations in the parameters of the first order
phase transition caused by including $\delta$ -meson exchange.
These results show how these variations will affect the integral
characteristics and structure of hybrid neutron stars with quark
matter cores.

\section{Lagrangian and thermodynamic characteristics of nucleonic systems}

The nonlinear Lagrangian density of an interacting multiparticle
system consisting of nucleons and isoscalar-scalar
$\sigma$-mesons, isoscalar-vector $\omega$-mesons,
isovector-scalar $\delta$-mesons, and isovector-vector
$\rho$-mesons has the following form in QHD:\footnote{We shall use
the natural system of units with $\hbar=c=1$.}
\begin{eqnarray}
\label{eq1} { \cal L}=\bar {\psi} _{N}\left[ \gamma ^{^{\mu
}}\left(
i\partial _{_{\mu }}-g_{\omega }\omega _{_{\mu }}(x)-\frac{1}{2}g_{\rho }%
\overrightarrow{\tau }_{_{N}}\overrightarrow{\rho }_{_{\mu
}}(x)\right) -\left( m_{_{N}}-g_{_{\sigma }}\sigma (x)-g_{_{\delta
}}\overrightarrow{\tau }_{_{N}}\overrightarrow{\delta }(x)\right)
\right] \psi _{N} \nonumber \\
+\frac{1}{2}\left( \partial _{_{\mu }}\sigma (x)\partial ^{^{\mu
}}\sigma (x)-m_{_{\sigma }}\sigma (x)^{2}\right) -U(\sigma(x))
+\frac{1}{2}m_{_{\omega }}^{2}\omega ^{^{\mu }}(x)\omega _{_{\mu
}}(x)-\frac{1}{4}\Omega _{_{\mu \nu}}(x)\Omega ^{^{\mu \nu }}(x)\nonumber \\%
+\frac{1}{2}\left( \partial _{_{\mu }}\overrightarrow{\delta
}(x)\partial
^{^{\mu }}\overrightarrow{\delta }(x)-m_{_{\delta }}^{2}\overrightarrow{%
\delta }(x)^{2}\right)+\frac{1}{2}m_{_{\rho }}^{2}\overrightarrow{\rho }^{^{\mu }}(x)%
\overrightarrow{\rho }_{_{\mu }}(x)-\frac{1}{4}\Re_{_{\mu \nu
}}(x)\Re^{^{\mu \nu}}\left( x\right)
\end{eqnarray}

\noindent where $x=x_{\mu}=(t,x,y,z)$, $\sigma(x)$, $\omega
_{_{\mu }}(x)$, $\overrightarrow{\delta }(x)$, and
$\overrightarrow{\rho }^{^{\mu }}(x)$ are the fields of the
$\sigma$, $\omega$, $\delta$, and $\rho$ exchange mesons,
respectively, $U(\sigma)$ is the nonlinear part of the potential
of the $\sigma$-field, given by \cite{Bog77}
\begin{equation}
\label{eq2}
U(\sigma)=\frac{b}{3}~m_{N}(g_{\sigma}\sigma)^{3}+\frac{c}{4}~(g_{\sigma}\sigma)^{4},
\end{equation}

\noindent $m_{N}$,~ $m_{\sigma}$,~ $m_{\omega}$, ~$m_{\delta}$,~
$m_{\rho}$ are the masses of the free particles,  $\psi _{N} =
\left({{\begin{array}{*{20}c} {\psi _{p}}  \hfill \\ {\psi _{n}}
\hfill \\\end{array}} } \right)$ is the isospin doublet for
nucleonic bispinors, and $\vec {\tau}$ are the isospin $2\times2$
Pauli matrices. The symbol $" {\overrightarrow{}} "$ denotes
vectors in isotopic spin space. This lagrangian, as in quantum
electrodynamics, also includes antisymmetric tensors of the vector
fields $\omega _{\mu }(x)$ and $\rho_{\mu }(x)$ given by
\begin{equation}
\label{eq3} \Omega _{\mu \nu}  \left( {x} \right) = \partial
_{\mu}  \omega _{\nu} \left( {x} \right) - \partial _{\nu}  \omega
_{\mu}  \left( {x} \right),\quad \;\Re _{\mu \nu}  \left( {x}
\right) = \partial _{\mu}  \rho _{\nu}  \left( {x} \right) -
\partial _{\nu}  \rho _{\mu}  \left( {x} \right).
\end{equation}

\noindent Here $g_{\sigma}  ,\;g_{\omega}  ,\;g_{\delta},  $ and
$g_{\rho}$ in (\ref{eq1}) denote the coupling constants of the
nucleon with the corresponding meson. In the RMF theory, the meson
fields $\sigma\left({x}\right)$, $\omega_{\mu}\left({x}\right)$,
$\vec{\delta}\left({x}\right)$ and $\vec{\rho}_{\mu}
\left({x}\right)$ are replaced by the (effective) fields $\bar
{\sigma} ,\;\,\bar{\omega} _{\mu} ,\;\,\bar {\vec {\delta
}},\;\bar {\vec {\rho} }_{\mu}$. Re-denoting the meson fields and
coupling constants according to
\begin{equation}
\label{eq4} g_{\sigma}  \bar {\sigma}  \equiv \sigma \,,\quad
g_{\omega}  \bar {\omega }_{0} \equiv \omega \,,\quad g_{\delta}
\bar {\delta} ^{\left( {3} \right)} \equiv \delta \,,\quad \quad
g_{\rho}  \bar {\rho} _{0}^{\left( {3} \right)} \equiv \rho \,,
\end{equation}
\begin{equation}
\label{eq5} (g_{\sigma}/m_{\sigma})^{2}\equiv a_{\sigma}, \quad
(g_{\omega}/m_{\omega})^{2}\equiv a_{\omega}, \quad
(g_{\delta}/m_{\delta})^{2}\equiv a_{\delta}, \quad
(g_{\rho}/m_{\rho})^{2}\equiv a_{\rho}\,
\end{equation}

\noindent
and introducing the asymmetry parameter
\begin{equation}
\label{eq6} \alpha = {{\left( {n_{n} - n_{p}}  \right)}
\mathord{\left/ {\vphantom {{\left( {n_{n} - n_{p}}  \right)}
{n}}} \right. \kern-\nulldelimiterspace} {n}}\;,
\end{equation}

\noindent the equations for the fields can be rewritten in the
form
\begin{equation}
\label{eq7} \sigma = a_{\sigma}  \left( {n_{s\,p} \left(
{n,\alpha}  \right) + n_{s\,n} \left( {n,\alpha}  \right)\, -
bm_{N} \sigma ^{2} - c\sigma ^{3}} \right),
\end{equation}
\begin{equation}
\label{eq8} \omega = a_{\omega}  n\,,
\end{equation}
\begin{equation}
\label{eq9} \delta = a_{\delta}  \left( {n_{s\,p} \left(
{n,\alpha}  \right) - n_{s\,n} \left( {n,\alpha}  \right)\,}
\right),
\end{equation}
\begin{equation}
\label{eq10} \rho = - \frac {1}{2} a_{\rho}n\alpha ,
\end{equation}

\noindent
where
\begin{equation}
\label{eq11} n_{s\,p} \left( {n,\alpha}  \right) = \frac{{1}}{{\pi
^{2}}}\int\limits_{0}^{k_{F\,} \left( {n} \right)\left( {1 -
\alpha} \right)^{1/3}} {\frac{{m_{p}^{\ast}  \left( {\sigma
,\delta} \right)}}{{\sqrt {k^{2} + m_{p}^{\ast}  \left( {\sigma
,\delta} \right)\,^{2}}} }} \;k^{2}dk\; \quad ,
\end{equation}
\begin{equation}
\label{eq12} n_{s\,n} \left( {n,\alpha}  \right) = \frac{{1}}{{\pi
^{2}}}\int\limits_{0}^{k_{F\,} \left( {n} \right)\left( {1 +
\alpha} \right)^{1/3}} {\frac{{m_{n}^{\ast}  \left( {\sigma
,\delta} \right)}}{{\sqrt {k^{2} + m_{n}^{\ast}  \left( {\sigma
,\delta} \right)\,^{2}}} }} \;k^{2}dk\; \quad ,
\end{equation}
\begin{equation}
\label{eq13} k_{F} \left( {n} \right) = \left( {\frac{{3\pi
^{2}n}}{{2}}} \right)^{1/3}\; \quad .
\end{equation}

The effective masses of the proton and neutron are determined by
the expressions
\begin{equation}
\label{eq14} m_{p}^{\ast}\left( {\sigma ,\delta}
\right)=m_{N}-\sigma-\delta, \quad m_{n}^{\ast}\left( {\sigma
,\delta} \right)=m_{N}-\sigma+\delta.
\end{equation}

If the constants $a_{\omega}$ and $a_{\rho}$ are known, equations
(\ref{eq8}) and (\ref{eq10}) determine the functions $\omega
\left( {n} \right)$ and $\rho \left( {n,\alpha} \right)$.
Moreover, a knowledge of the other constants $a_{\sigma}$,
$a_{\delta}$, $b$, and $c$ makes it possible to solve the set of
equations (\ref{eq7}), (\ref{eq9}), (\ref{eq11}), (\ref{eq12}) in
a self-consistent way and to determine the remaining two meson
field functions $\sigma \left( {n,\alpha} \right)$ and $\delta
\left( {n,\alpha} \right)$.

The energy density of the nuclear $np$ matter as a function of the
concentration $n$ and the asymmetry parameter $\alpha $ has the
form
\begin{eqnarray}{}\label{eq15}
\varepsilon ({n,\alpha}) =
\frac{{1}}{{\pi^{2}}}\int\limits_{0}^{k_{F\,} \left( {n}
\right)\left( {1 - \alpha} \right)^{1/3}} {\sqrt {k^{2} + \left(
{m_{N} - \sigma - \delta}  \right)^{\,2}}}~  k^{2}dk \nonumber\\ +
\frac{{1}}{{\pi ^{2}}}\int\limits_{0}^{k_{F\,} \left( {n}
\right)\left( {1 + \alpha}  \right)^{1/3}} {\sqrt {k^{2} + \left(
{m_{N} - \sigma + \delta}  \right)^{\,2}}}~  k^{2}dk + \tilde
{U}\left( {\sigma} \right) + \frac{{1}}{{2}}\left(
{\,\frac{{\sigma ^{\,2}}}{{a_{\sigma} } } + \,\frac{{\omega
^{2}}}{{a_{\omega} } } + \frac{{\delta ^{\,2}}}{{a_{\delta} } } +
\frac{{\rho ^{\,2}}}{{a_{\rho} } }}\right),
\end{eqnarray}

\noindent where
\begin{equation}
\label{eq16} \tilde {U}\left( {\sigma}  \right) =
\frac{{b}}{{3}}\,m_{N} \,\sigma ^{3} + \frac{{c}}{{4}}\,\sigma
^{4}\;.
\end{equation}

For the pressure of the nuclear matter we obtain
\begin{eqnarray}{}
 P({n,\alpha}) = \frac{1}{\pi
^{2}}\int\limits_{0}^{k_{F\,} \left( {n} \right)\left( {1 -
\alpha} \right)^{1/3}} {\left( {\sqrt {k_{F} \left( {n}
\right)^{2}\left( {1 - \alpha} \right)^{2/3} + \left( {m_{N} -
\sigma - \delta}  \right)^{\,2}} - \sqrt {k^{2} + \left( {m_{N} -
\sigma - \delta}  \right)^{2}}}  \right)\;}k^{2}dk  \nonumber \\
+\frac{{1}}{{\pi ^{2}}}\int\limits_{0}^{k_{F\,} \left( {n}
\right)\left( {1 + \alpha}  \right)^{1/3}} {\left({\sqrt {k_{F}
\left( {n} \right)^{2}\left( {1 + \alpha} \right)^{2/3} + \left(
{m_{N} - \sigma + \delta}  \right)^{\,2}} - \sqrt {k^{2} + \left(
{m_{N}-\sigma + \delta}\right)^{2}}}  \right)}k^{2}dk \nonumber
\end{eqnarray}
\begin{eqnarray}{}
\label{eq17} - \tilde {U}\left( {\sigma}  \right) +
\frac{{1}}{{2}}\left( { - \frac{{\sigma ^{\,2}}}{{a_{\sigma} } } +
\,\frac{{\omega ^{2}}}{{a_{\omega }} } - \frac{{\delta
^{\,2}}}{{a_{\delta} } } + \frac{{\rho
^{\,2}}}{{a_{\rho}}}}\right).
\end{eqnarray}

The chemical potentials of the proton and neutron are given by
\begin{eqnarray}
\label{eq18} \mu_{p}(n,\alpha)=\sqrt {k_{F} \left( {n}
\right)^{2}\left( {1 - \alpha} \right)^{2/3} + \left( {m_{N} -
\sigma - \delta}
\right)^{\,2}} + \omega + \frac{1}{2}\rho,\nonumber\\
\mu_{n}(n,\alpha)=\sqrt {k_{F} \left( {n} \right)^{2}\left( {1 +
\alpha} \right)^{2/3} + \left( {m_{N} - \sigma + \delta}
\right)^{\,2}}+ \omega - \frac{1}{2}\rho.
\end{eqnarray}

\bigskip

\section{Determination of the constants for the model}
\subsection{Empirical characteristics of saturated nuclear matter and the
constants of the theory}

In order to determine the constants for the theory, $a_{\sigma}
\,,\;\,a_{\omega} \,,\;\,a_{\delta} \,,\,\,a_{\rho} \,,\;\,b$, and
$c$ we can derive a system of equations relating these parameters
to known empirical characteristics of symmetric nuclear matter at
the saturation concentration $n_{0}$ \cite{Gl00}. Given that the
effective mass of a nucleon in symmetric nuclear matter ($\alpha $
= 0) at the saturation concentration $ n_{0}$ is related to the
bare nucleon mass by
\begin{equation}
\label{eq19} m_{N}^{\ast}  = \gamma \,m_{N} \,,
\end{equation}

\noindent where $\gamma $ is a constant between $0.7$  and $0.8$,
for the $\sigma $ field at the saturation concentration $ n_{0}$
we have
\begin{equation}
\label{eq20} \sigma _{0} = \left( {1 - \gamma}  \right)m_{N} .
\end{equation}

Equations (\ref{eq9}) and (\ref{eq10}) imply that $\delta _{0} =
0$ and $\rho _{0} = 0$ at the saturation concentration in
saturated nuclear matter. Given the requirement that the energy $
\varepsilon \left( {n,\alpha} \right)/n $ per nucleon should have
a minimum at $ n = n_{0}$ and $\alpha = 0$, we obtain
\begin{equation}
\label{eq21} \left. {\frac{{d\varepsilon \left( {n,\alpha}
\right)}}{{dn}}} \right|_{
\begin{array}{l}
{n = n_{0}}  \\
{\alpha = 0} \\
\end{array}} = \frac{\varepsilon \left( {n_{0},0}\right)}{n_{0}} = m_{N}+f_{0},
\end{equation}

\noindent where $f_{0} = B/A $ is the specific binding energy of
the nucleus, neglecting the Coulomb interaction and finite size
effects of the nucleus.

Using Eq. (\ref{eq15}), Eq. (\ref{eq21}) yields
\begin{equation}
\label{eq22} a_{\omega}  = \frac{{1}}{{n_{0}} }\left( {m_{N} +
f_{0} - \sqrt {k_{F} \left( {n_{0}}  \right)^{2} + \left( {m_{N} -
\sigma _{0}}  \right)^{\,2}}} \right)\;.
\end{equation}

The $\omega_{0}$ field for symmetric matter at $n_{0} $, on the
other hand, is given by
\begin{equation}
\label{eq23} \omega _{0} = a_{\omega}  n_{0} = m_{N} + f_{0} -
\sqrt {k_{F} \left( {n_{0} } \right)^{2} + \left( {m_{N} - \sigma
_{0}}  \right)^{\,2}}\; .
\end{equation}

From the equation for the $\sigma$ field (\ref{eq7}), we have
 \begin{equation}
\label{eq24} \frac{\sigma_{0}}{a_{\sigma}} = \frac{{2}}{{\pi
^{2}}}\int\limits_{0}^{k_{F} \left( {n_{0}} \right)} {\frac{\left(
{m_{N} - \sigma_{0}}\right)} {\sqrt {k^{2} + \left( {m_{N} -
\sigma_{0}}\right)\,^{2}}} } \;k^{2}dk -
bm_{N}\sigma_{0}^{2}-c\sigma_{0}^{3}\;.
\end{equation}

The energy density $\varepsilon _{0} = n_{0} \left( {m_{N} +
f_{0}} \right)\;$ for saturated nuclear matter at the saturation
concentration $n_{0} $ can be written in the form
\begin{equation}
\label{eq25} \varepsilon _{0} = \frac{{2}}{{\pi
^{2}}}\int\limits_{0}^{k_{F\,} \left( {n_{0}}  \right)} {\sqrt
{k^{2} + \left( {m_{N} - \sigma _{0}} \right)^{\,2}}}  k^{2}dk +
\frac{{b}}{{3}}\,m_{N} \,\sigma _{0} ^{3} +
\frac{{c}}{{4}}\,\sigma _{0} ^{4}\; + \frac{{1}}{{2}}\left(
{\frac{{\sigma _{0} ^{\,2}}}{{a_{\sigma} } } + \,n_{0}
^{2}a_{\omega} }  \right)\;.
\end{equation}

An important empirical characteristic which, in a certain way,
couples the phenomenological constants of the theory, is the
modulus of compression of the nuclear matter, which is defined as
\begin{equation}
\label{eq26} K = \left. {9\;n_{0}
^{2}\frac{{d^{2}}}{{dn^{2}}}\left( {\frac{{\varepsilon \left(
{n,\alpha}  \right)}}{{n}}} \right)} \right|_{\begin{array}{l}
 {n = n_{0}}  \\
 {\alpha = 0} \\
 \end{array}} .
\end{equation}

Substituting Eq. (\ref{eq15}) in Eq. (\ref{eq26}) gives
\begin{eqnarray}
\label{eq27}
 K = 9a_{\omega}  n_{0} + 3\frac{{k_{F} \left( {n_{0}}  \right)^{2}}}{{\sqrt
{k_{F} \left( {n_{0}}  \right)^{2} + \left( {m_{N} - \sigma _{0}}
\right)^{\,2}}} } \nonumber \\
 \;\quad \; - 9\frac{{n_{0} \left( {m_{N} - \sigma _{0}}
\right)^{2}}}{{k_{F} \left( {n_{0}}  \right)^{2} + \left( {m_{N} -
\sigma _{0}}  \right)^{\,2}}}\frac{{1}}{{\frac{{1}}{{a_{\sigma} }
} + \frac{{2}}{{\pi ^{2}}}\int\limits_{0}^{k_{F\,} \left( {n_{0}}
\right)} {\frac{{k^{4}dk}}{{\left[ {k^{2} + \left( {m_{N} - \sigma
_{0}} \right)^{2}} \right]^{3/2}}}} \; + 2b\,m_{N} \,\sigma _{0} +
3c\,\sigma _{0} ^{2}}}\;.
\end{eqnarray}

In the semi-empirical Weizs\"{a}cker formula the term for the
specific energy of the asymmetry of the nucleonic system is given
by
\begin{equation}
\label{eq28} \frac{{\varepsilon _{sym}} }{{n}} = E_{sym} \left(
{n} \right)\,\alpha ^{2}.
\end{equation}

The coefficient of asymmetry energy,  $E_{sym} \left( {n}
\right)$, is defined as
\begin{equation}
\label{eq29} E_{sym} \left( {n} \right) = \left.
{\frac{{1}}{{2n}}\frac{{d^{2}\varepsilon \left( {n,\alpha}
\right)}}{{d\alpha ^{2}}}} \right|_{\alpha = 0} .
\end{equation}

Using Eq. (\ref{eq15}), we obtain the following expression for the
symmetry energy at the saturation concentration of the nuclear
matter, $E_{sym}^{\left( {0} \right)} = E_{sym} \left( {n_{0}}
\right)$
\begin{eqnarray}
\label{eq30}
 E_{sym}^{\left( {0} \right)} = \frac{{n_{0}} }{{8}}a_{\rho}  + \frac{{k_{F}
\left( {n_{0}}  \right)^{2}}}{{6\sqrt {k_{F} \left( {n_{0}}
\right)^{2} +
\left( {m_{N} - \sigma _{0}}  \right)^{\,2}}} } \nonumber \\
 \quad \quad - \frac{{1}}{{2}}\frac{{n_{0} \left( {m_{N} - \sigma _{0}}
\right)^{2}}}{{k_{F} \left( {n_{0}}  \right)^{2} + \left( {m_{N} -
\sigma _{0}}  \right)^{\,2}}}\frac{{1}}{{\frac{{1}}{{a_{\delta} }
} + \frac{{2}}{{\pi ^{2}}}\int\limits_{0}^{k_{F\,} \left( {n_{0}}
\right)} {\frac{{k^{4}dk}}{{\left( {k^{2} + \left( {m_{N} - \sigma
_{0}} \right)^{2}} \right)^{3/2}}}}} }\;.
\end{eqnarray}

\bigskip

\subsection{Numerical determination of the constants for the theory}
In order to determine the constants for the theory we have used
the following values of the known nuclear parameters at
saturation: $m_{N} = 938,93$ MeV, $\gamma = m_{N}^{\ast}/m_{N} =
0,78$, saturation concentration of nuclear matter $n_{0} = 0,153$
fm$^{-3}$, specific binding energy $f_{0} = - 16,3$ MeV, modulus
of compression $ K=300$ MeV, and $E_{sym}^{\left( {0} \right)} =
32,5$ MeV. Equations (\ref{eq20}) and (\ref{eq23}) can be used to
determine the $\sigma _{0} $ and $\omega _{0} $ fields. Then Eqs.
(\ref{eq22}), (\ref{eq24}), (\ref{eq25}), (\ref{eq27}) and
(\ref{eq30}) form a system of five equations for the six unknown
constants, $a_{\sigma} \,,\;\,a_{\omega} \,,\;\,a_{\delta}
\,,\,\,a_{\rho}  \,,\;\,b$ and $ c $. It can be seen from Eq.
(\ref{eq30}) that including the interaction channel involving the
isovector-scalar $\delta $ -meson leads to a certain correlation
between the values of $\;\,a_{\delta}  \,$ and $\,a_{\rho} \,$.

\begin{table}[h]
\caption{Values of the Constant $a_{\rho}$ for Different Values of
$a_{\delta}$}
 \label{1} \centering
\begin{tabular}{|c|c|c|c|c|c|c|c|}
\hline  $a_{\delta }=(g_{\delta }/m_{\delta })^{2},\; fm^{2}$ &
$0$ & $0.5$ & $1$ & $1.5$ & $2$ & $2.5$ & $3$ \\
\hline $a_{\rho}=(g_{\rho }/m_{\rho })^{2},\; fm^{2}$ & $4.794$ &
$6.569$ & $8.340$ & $10.104$ & $11.865$ & $13.621$ & $15.372$ \\
\hline
\end{tabular}%
\end{table}

Table 1 lists the values of $a_{\rho}  $ for various values of the
constant $ a_{\delta}$. In order to clarify the role of the
$\delta $ -meson, in the following we shall take $a_{\delta}  =
2.5$~fm$^{2} $ \cite{Liu02}. The absence of the $\delta$
interaction channel will correspond to an interaction constant $
a_{\delta}  = 0$. Note that the value used here, $a_{\delta}  =
2.5$~fm$^{2}$, is in good agreement with Ref. 18, where a
microscopic Dirac-Bruckner-Hartree-Fock theory is applied to
asymmetric nuclear matter and exotic nuclei in a study of the
density dependence of the meson-nucleon coupling constants.
According a plot of $a_{\delta}$ as a function of concentration
$n$ in Fig. 2 of Ref. \cite{Hof01}, the average value of
$a_{\delta} $ in the range $n \approx ~ $0.1 $ \div $
0.3~fm$^{-3}$ is on the order of 65 GeV$^{-2} \approx $
2.5~fm$^{2}$.

\begin{table}[h]
\centering \caption{Constants for the Theory without ($\sigma
\omega \rho $) and with ($\sigma \omega \rho \delta$) a $\delta$
-Meson Field} \label{2}
\begin{tabular}{|c|c|c|c|c|c|c|}
\hline & $a_{\sigma }$,\ fm$^{2}$ & $a_{\omega }$,\ fm$^{2}$ &
$a_{\delta }$,\ fm$^{2}$ & $a_{\rho }$,\ fm$^{2}$ & $b$,\
fm$^{-1}$ & $c$ \\ \hline $\sigma \omega \rho $ & $9.154$ &
$4.828$ & $0$ & $4.794$ & $1.654\cdot 10^{-2}$ & $1.319\cdot
10^{-2}$ \\ \hline
$\sigma \omega \rho \delta $ & $9.154$ & $4.828$ & $2.5$ & $13.621$ & $%
1.654\cdot 10^{-2}$ & $1.319\cdot 10^{-2}$ \\ \hline
\end{tabular}%
\end{table}
Table 2 lists the values of the parameters obtained by numerical
solution of the system of five Eqs. (\ref{eq22}), (\ref{eq24}),
(\ref{eq25}), (\ref{eq27}) and (\ref{eq30}) without ($\sigma
\omega \rho $) and with ($\sigma \omega \rho \delta $) the
isovector-scalar $\delta$ -meson interaction channel.

\bigskip

\section{Characteristics of $\beta$ -equilibrium $npe$ plasmas and the equation of state of neutron star matter in the nucleonic
phase } The values of the constants $a_{\sigma}, \;a_{\omega},
\;a_{\delta}, \;a_{\rho}, \;b$, and $c$ for the relativistic
mean-field theory obtained in the previous section (See Table 2.)
can be used to calculate various characteristics of matter with an
asymmetric proton - neutron composition ($np$ -matter) and of
$\beta$ -equilibrium $npe$ -matter. In terms of the relativistic
mean-field theory, the lagrangian density for the $npe$ -plasma is
\begin{equation}
\label{eq31} {\cal L}_{NM} = {\cal L} + \bar {\psi} _{e} \,\left(
{i\gamma ^{\mu }\partial _{\mu}  - m_{e}}  \right)\,\psi _{e}
\quad ,
\end{equation}

\noindent where ${\cal L}$ is the lagrangian of the system
consisting of nucleons and $\sigma \,\omega \,\rho\, \delta$
mesons (See Eq. (\ref{eq1})), $\,\psi _{e} $ is the electron wave
function, and $ m_{e}$ is the electron mass. In this case, we find
the energy density of the $npe$ -plasma to be
\begin{equation}
\label{eq32} \varepsilon _{NM} \left( {n,\alpha ,\mu _{e}}
\right) = \varepsilon \left( {n,\alpha}  \right) + \;\varepsilon
_{e} \left( {\mu _{e}}  \right),
\end{equation}

\noindent where $\varepsilon \left( {n,\alpha}  \right)$ is the
energy density of the $n\,p\,\sigma \,\omega \,\rho\, \delta $
system defined by Eq. (\ref{eq26}),
\begin{equation}
\label{eq33} \varepsilon _{e}(\mu_{e})= \frac{1}{\pi^{2}}\int
\limits_{0}^{\sqrt{\mu_{e}^{2}-m_{e}^{2}}} \sqrt {k^{2} +
m_{e}^{2}} ~ k^{2}dk
\end{equation}

\noindent is the contribution of the electrons to the energy
density, and $\mu _{e} $ is the chemical potential of the
electrons.
For the pressure of the $npe$ -plasma, we have
\begin{equation}
\label{eq34} P_{NM} \left( {n,\alpha ,\mu _{e}}  \right) =
\,P\left( {n,\alpha}  \right) + \frac{{1}}{{3\pi ^{2}}}\mu _{e}
\left( {\mu _{e} ^{2} - m_{e} ^{2}} \right)^{3/2} - \varepsilon
_{e} \left( {\mu _{e}}  \right) \quad .
\end{equation}

Depending on the coefficient of surface tension, $\sigma _{s} $,
it is known that a phase transition of nuclear matter into quark
matter can occur in two ways \cite{Hei93}. It can either have the
character of an ordinary first order phase transition with a
discontinuous density change (Maxwell's rule) or mixed
nucleon-quark matter can be formed with a continuous variation in
the pressure and density \cite{Gl92}. In the second case, the
condition of global electrical neutrality implies that, in order
to determine the parameters of the phase transition and the
equation of state of the mixed phase, it is necessary to know the
equation of state of the $ \beta$ -equilibrium charged $npe$
-plasma. To find the characteristics of the $\beta$ -equilibrium,
but not necessarily the neutral, $npe$ -plasma, one has to solve
the system of four equations (7)-(10) for specified values of the
concentration $n$ and asymmetry parameters $\alpha$, and find the
unknown mean meson fields $\sigma \left( {n,\alpha} \right)$,
$\,\omega \left( {n} \right)$, $\delta \left( {n,\alpha} \right)$
è $\rho \left( {n,\alpha} \right)$. Equations (18) can be used to
determine the chemical potentials of the nucleons, $\mu_{n} \left(
{n,\alpha}  \right)$ and $\mu _{p} \left( {n,\alpha} \right)$, so
that, using the $\beta $ -equilibrium condition, it is possible to
find the electron chemical potential, and, ultimately, the energy
density $\varepsilon _{NM}$ and pressure $P_{NM}$  of the $\beta$
-equilibrium $npe$-plasma.
\begin{equation}
\label{eq35} \mu _{e} \left( {n,\alpha}  \right) = \mu _{n} \left(
{n,\alpha}  \right) - \mu _{p} \left( {n,\alpha}  \right)
\end{equation}

Figure 1 is a three dimensional plot of the energy per baryon,
$E_{b} \left( {n,\alpha}  \right) = {{\varepsilon _{NM}}
\mathord{\left/ {\vphantom {{\varepsilon _{NM}} {n}}} \right.
\kern-\nulldelimiterspace} {n}}$, as a function of the
concentration $n$ and asymmetry parameter $\alpha$ for the case of
a $\beta $ -equilibrium charged $npe$ -plasma. The lines
correspond to different fixed values of the charge per baryon, $q
= {{\left( {n_{p} - n_{e}} \right)} \mathord{\left/ {\vphantom
{{\left( {n_{p} - n_{e}} \right)} {n}}} \right.
\kern-\nulldelimiterspace} {n}} = {{\left( {1 - \alpha} \right)}
\mathord{\left/ {\vphantom {{\left( {1 - \alpha} \right)} {2}}}
\right. \kern-\nulldelimiterspace} {2}} - {{n_{e}} \mathord{\left/
{\vphantom {{n_{e}}  {n}}} \right. \kern-\nulldelimiterspace}
{n}}$.
\begin{figure} [h]
\begin{center}
  \begin{minipage}[t]{0.48\linewidth}
   \begin{center}
   \includegraphics[height=0.9\linewidth, width=0.9\linewidth]%
    {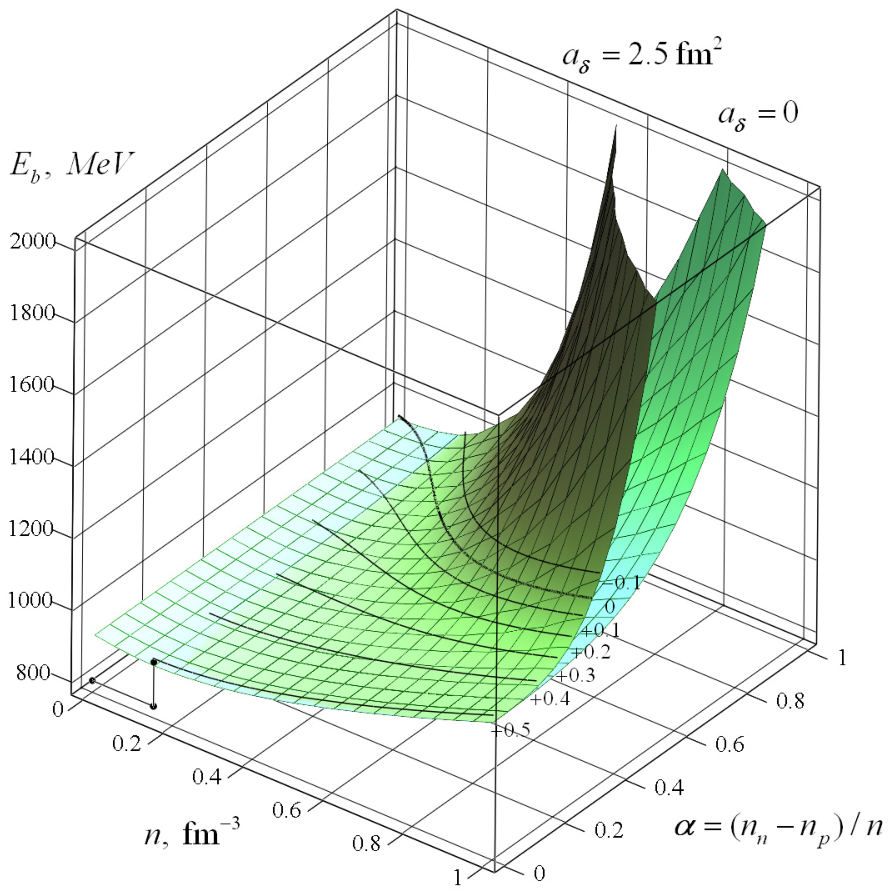}
    \caption {\small{Three dimensional representation of the energy per
baryon $E_{b}$ as a function of the baryon number density $n$ and
the asymmetry parameter $\alpha$ in the case of a $\beta$
-equilibrium charged $ npe$ -plasma. The upper surface corresponds
to the "$\sigma \omega \rho \delta $" model, and the lower, to
"$\sigma \omega \rho $". The lines correspond to different values
of the charge per baryon. }}
    \end{center}
  \end{minipage}\label{Fig1} \hfil\hfil
  \begin{minipage}[t]{0.48\linewidth}
   \begin{center}
   \includegraphics[height=0.9\linewidth, width=0.9\linewidth]%
    {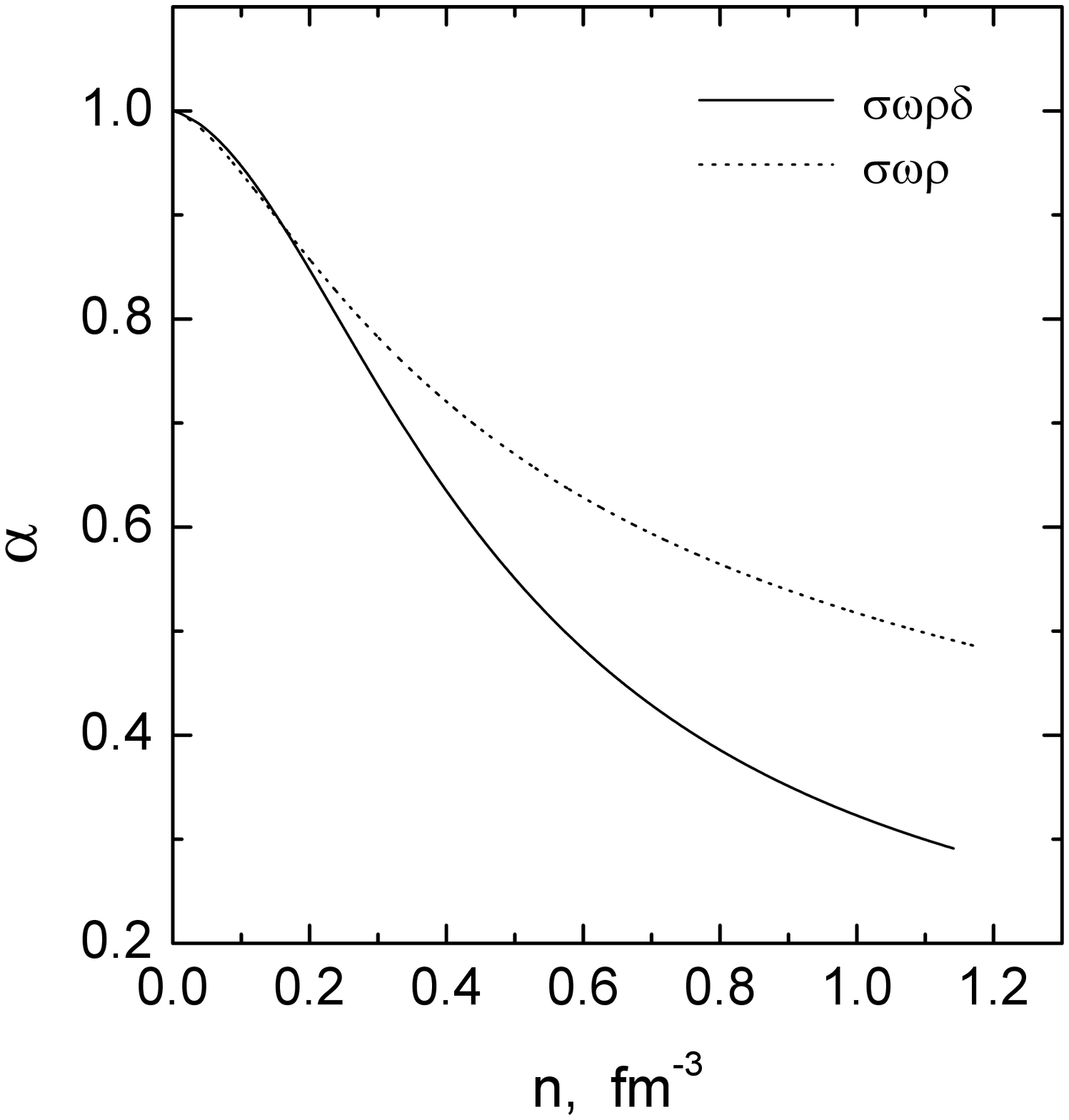}
    \caption {\small{The asymmetry parameter as a function of the
     concentration $n$ for a $\beta$ -equilibrium, uncharged $npe$
    -plasma. The smooth curve corresponds to the "$\sigma \omega \rho
    \delta$" model and the dashed curve, to "$\sigma \omega \rho $".
     }}
    \end{center}
  \end{minipage}\label{Fig2}
 \end{center}
\end{figure}

The thick line corresponds to $\beta$ -equilibrium electrically
neutral $npe$ -matter. The lower surface corresponds to the
"$\sigma \omega \rho $" model and the upper, to the "$\sigma
\omega \rho \delta $" model. Clearly, including a $\delta$ -meson
field increases the energy per nucleon, and the change is greater
for larger values of the asymmetry parameter of the nuclear
matter. For a fixed value of the specific charge, the asymmetry
parameter falls off monotonically as the concentration is
increased. Figure 2 is a plot of the asymmetry parameter as a
function of the concentration $n$ for the case of an electrically
neutral "$npe$" -plasma for the "$\sigma \omega \rho $" and
"$\sigma \omega \rho \delta $" models. It can be seen that
including a $\delta$ -meson field for a fixed concentration $n$
will reduce the asymmetry parameter $\alpha$.

\begin{figure} [h]
\begin{center}
  \begin{minipage}[t]{0.48\linewidth}
   \begin{center}
   \includegraphics[height=0.9\linewidth, width=0.9\linewidth]%
    {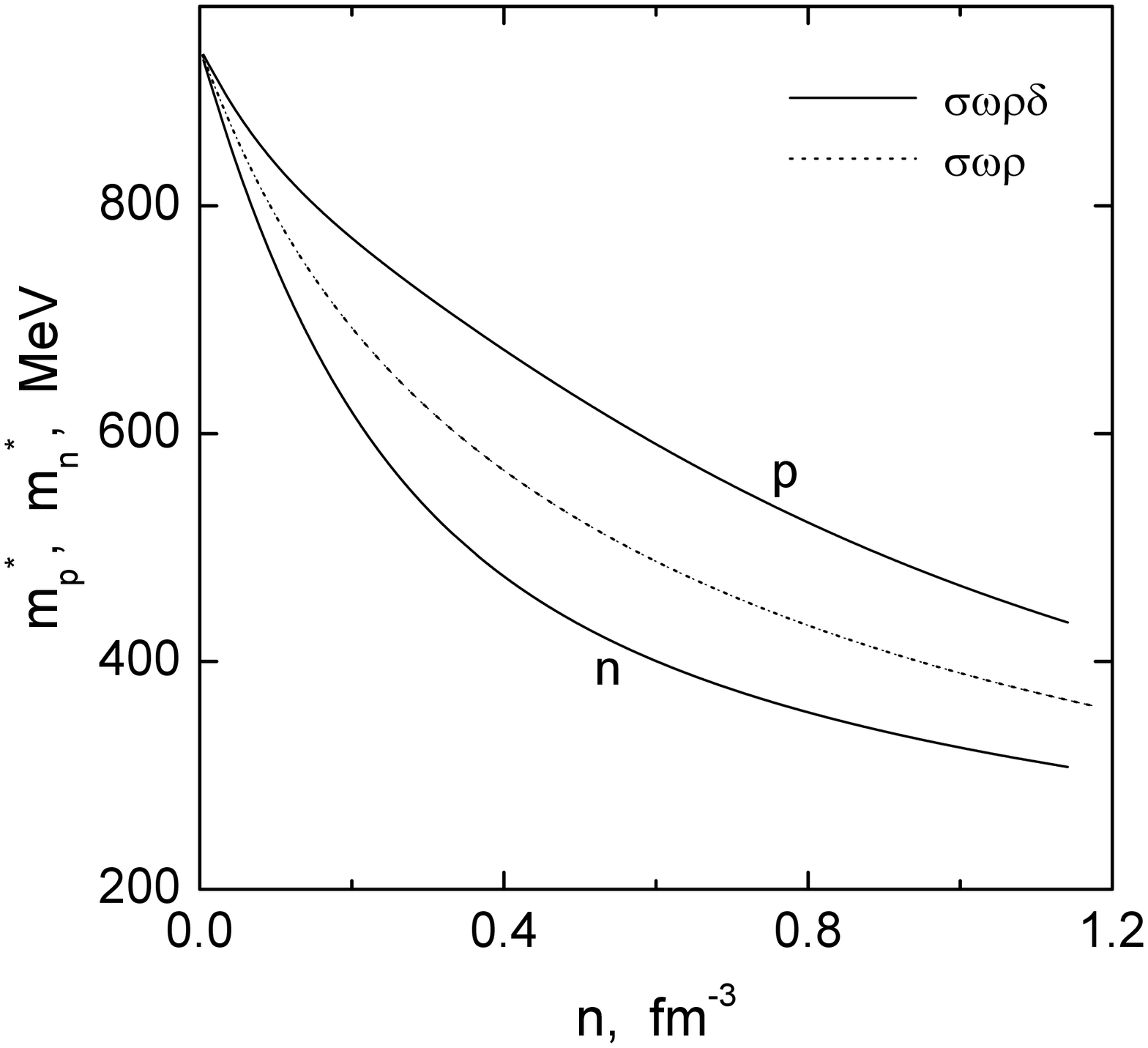}
    \caption {\small{ The effective nucleon masses as functions of the baryon
concentration $n$ for a $\beta$ -equilibrium, uncharged $npe$
-plasma for the "$\sigma \omega \rho \delta $" model. The dashed
curve corresponds to the "$\sigma \omega \rho$" model. }}
    \end{center}
  \end{minipage}\label{Fig3} \hfil\hfil
  \begin{minipage}[t]{0.48\linewidth}
   \begin{center}
   \includegraphics[height=0.9\linewidth, width=0.9\linewidth]%
    {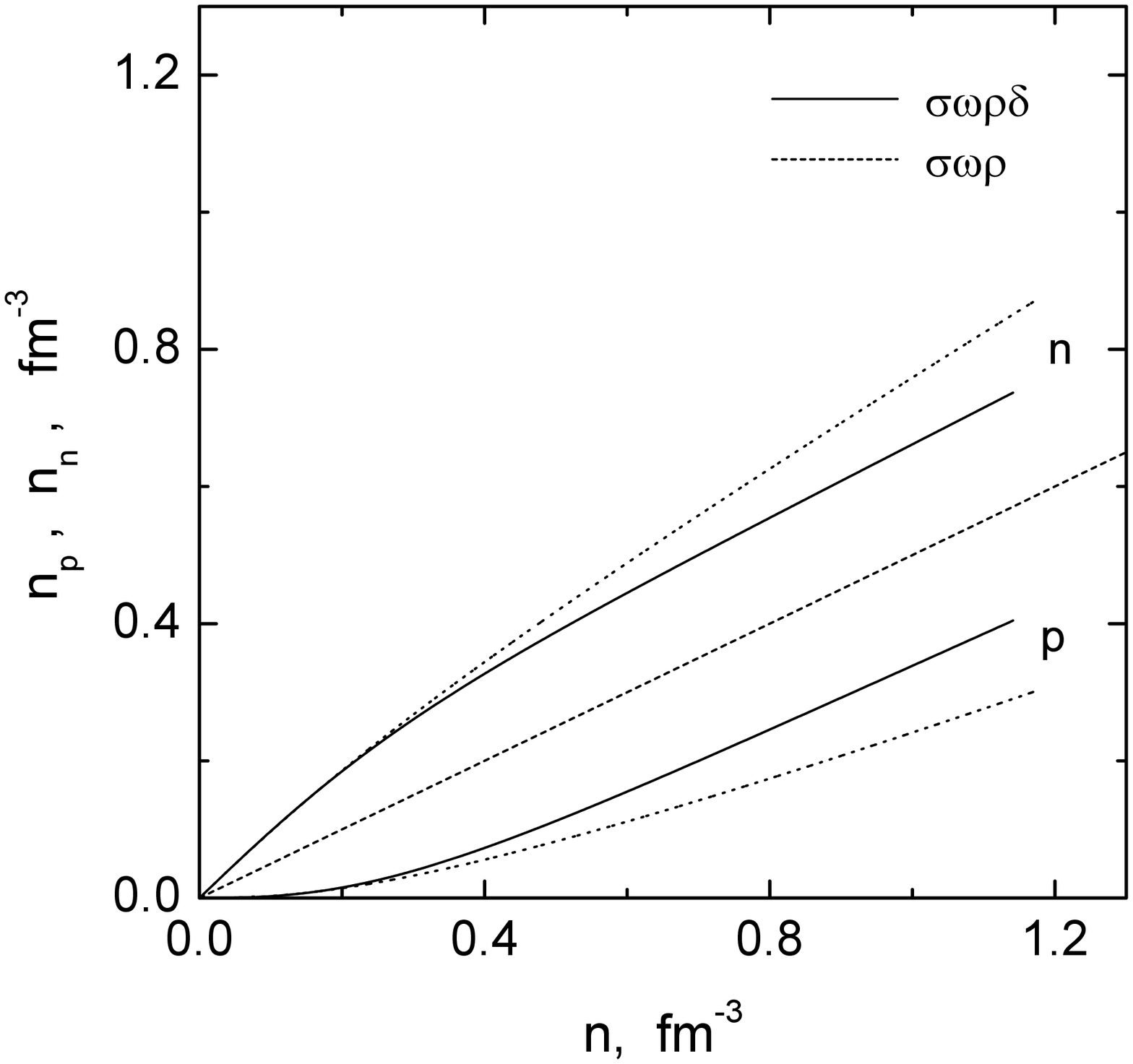}
    \caption {\small{The concentrations of protons and neutrons as functions of
the baryon concentration $n$ for a  $\beta$ -equilibrium,
uncharged $npe$ -plasma. The smooth curves correspond to the
"$\sigma \omega \rho \delta $" model and the dotted curves, to the
"$\sigma \omega \rho$" model. The dashed (straight) line
corresponds to isospin symmetric matter.}}
    \end{center}
  \end{minipage}\label{Fig4}
 \end{center}
\end{figure}

Figure 3 shows the effective masses of the protons and neutrons in
a $\beta$ -equilibrium uncharged $npe$ -plasma as functions of the
baryon concentration $n$ for the "$\sigma \omega \rho \delta $"
model. Note that the effective masses of the protons and neutrons
are the same in the "$\sigma \omega \rho$" model. Including the
-meson mean field breaks the symmetry, in this sense, between the
protons and the neutrons; the effective mass of the protons in
this kind of medium is greater than that of the neutrons, i.e.,
there is a split in the values of the effective masses for the
protons and neutrons.

Figure 4 contains plots of the concentrations of the protons and
neutrons as functions of the baryon concentration $n$ for a
$\beta$-equilibrium uncharged $npe$ -plasma. The dashed (straight)
line corresponds to the case of isospinsymmetric matter. It is
clear from this figure that the presence of a $\delta$ -meson
field reduces the neutron concentration and increases that of the
protons.

Our calculated equation of state for electrically neutral $\beta$
-equilibrium $npe$ -matter (the neutron star matter in nucleonic
phase) using the "$\sigma \omega \rho \delta $" model is shown in
Fig. 5. Our equation of state (the segment of the curve labelled
"MFT-$\sigma \omega \rho \delta $") has been matched to the
Baym-Bethe-Pethick (BBP) equation of state [21] in the region of
normal nuclear densities. The Malone-Bethe-Johnson (MBJ) equation
of state \cite{MBJ75} is also shown for comparison.

\begin{center}
\begin{figure}[h]
 \begin{minipage}[c]{0.48\linewidth}
   \begin{center}
   \includegraphics[height=0.9\linewidth, width=0.9\linewidth]%
    {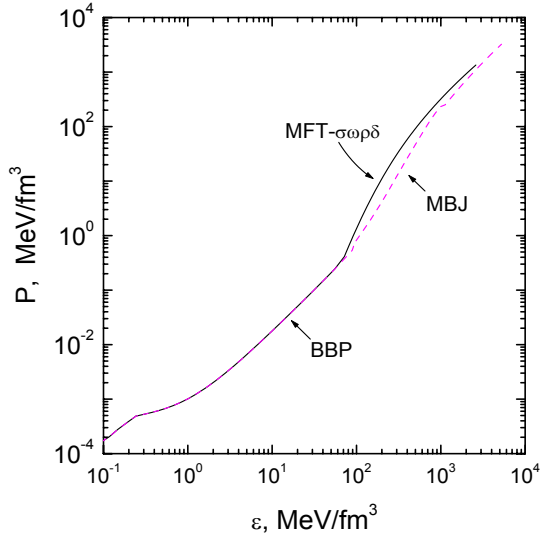}
    \caption {\small{ The equation of state for neutron star matter in the
nucleonic phase. The segment "MFT-$\sigma \omega \rho \delta$"
represents the results of this paper and "MBJ," those of Ref.
\cite{MBJ75}. The region corresponding to nuclear - neutron (Aen)
matter is described by the BBP equation of state \cite{BBP71}. }}
   \end{center}
  \end{minipage}\label{Fig5}
\end{figure}
\end{center}

\bigskip
\section{The equation of state for a quark-electron ("$udse$") plasma}
In order to describe the quark phase we have used an improved
version of the MIT bag model \cite{Chod74}, in which the
interactions between the $u,\;d$, and $s$ quarks inside the bag
are accounted for in a one-gluon exchange approximation
\cite{Far84}. The quark phase consists of three quark flavors,
$u,\;d$, and $s$ and electrons that are in equilibrium with
respect to weak interactions via the reactions \textbf{}
\[
d \to u + e^{ -}  + \tilde {\nu} _{e} \;,\quad \;\;u + e^{ -}  \to
d + \nu _{e} \;\;,\quad \;s \to u + e^{ -}  + \tilde {\nu} _{e}
\;,\quad \;\;u + e^{ -}  \to s + \nu _{e} \;\;\;.
\]

Since the $\nu _{e} $ and $\tilde {\nu} _{e}$ particles leave the
system, the energy of the system decreases and the reactions with
neutrino emission continue until the condition $\mu _{\nu} = 0$
holds for the chemical potential of the neutrinos. Then, the
following conditions hold for the chemical potentials of the
$u,\;d,\;s$, and $e$ particles:
\begin{equation}
\label{eq36} \mu _{d} = \mu _{s} \equiv \mu \quad , \quad \mu _{u}
+ \mu _{e} = \mu \quad .
\end{equation}

The following expression for the density of the thermodynamic
potential $\Omega_{f}$ of the quark flavor $f$ ($f = u,d,s$) has
been obtained in a framework of quantum hadrodynamics (QHD)
\cite{Far84}:
\begin{eqnarray}
\label{eq37} \Omega _{f} \left( {\mu _{f}}  \right) = -
\frac{{1}}{{4\pi ^{2}}}\left\{ {\mu _{f} \sqrt {\mu _{f} ^{2} -
m_{f} ^{2}} \left( {\mu _{f} ^{2} - \frac{{5}}{{2}}m_{f} ^{2}}
\right) + \frac{{3}}{{2}}m_{f} ^{4}\; ln\left( {\frac{{\mu _{f} +
\sqrt {\mu _{f}^{2} - m_{f}^{2}} } }{{m_{f}} }} \right)}
\right.\nonumber \\
 - 2\frac{{\alpha _{s}} }{{\pi} }\left[ {3\left( {\mu _{f} \sqrt
{\mu _{f}^{2} - m_{f}^{2}}  - m_{f}^{2} \; ln\frac{{\mu _{f} +
\sqrt {\mu _{f}^{2} - m_{f}^{2}} } }{{\mu _{f}} }} \right)}
\right.^{2} - 2\left( {\mu _{f}^{2} - m_{f}^{2}}  \right)^{2}
\nonumber\\  - \left. { 3m_{f}^{4} \; ln^{2}\left( {\frac{{m_{f}}
}{{\mu _{f}} }} \right) + 6m_{f} ^{2} \; ln\left( {\frac{{\tilde
{\rho} }}{{\mu _{f}} }} \right)\left. {\left( {\mu _{f} \sqrt {\mu
_{f}^{2} - m_{f}^{2}} - m_{f}^{2}\; ln\frac{{\mu _{f} + \sqrt {\mu
_{f}^{2} - m_{f}^{2}} } }{{m_{f}} }} \right)} \right]} \right\}
 ,
\end{eqnarray}

\noindent where $\alpha _{s} = g^{2}/4\pi$, $g$ is the QHD
coupling constant, and $\tilde {\rho} \approx m/3 \approx 313$ MeV
is the renormalization parameter. The quark concentrations are
given by
\begin{equation}
\label{eq38} n_{f} \left( {\mu _{f}}  \right) = \frac{{\mu _{f}
^{2} - m_{f}^{2}} }{{\pi ^{2}}}\left\{ {\;\sqrt {\mu _{f} ^{2} -
m_{f} ^{2}} - 2\frac{{\alpha _{s} }}{{\pi} }\left[ {\mu _{f} -
\frac{{3m_{f}^{2}} }{{\sqrt {\mu _{f} ^{2} - m_{f} ^{2}}} }\;
ln\frac{{\mu _{f} + \sqrt {\mu _{f} ^{2} - m_{f} ^{2}} }}{{\tilde
{\rho} }}} \right]} \right\}\;.\;
\end{equation}

\noindent The thermodynamic potential $\Omega _{e}$ and
concentration of the electrons are given by
\begin{equation}
\label{eq39} \Omega _{e} \left( {\mu _{e}}  \right) = -
\frac{{1}}{{\pi ^{2}}}\int\limits_{0}^{\sqrt {\mu _{e} ^{2} -
m_{e} ^{2}}}  {\left( {\mu _{e} - \sqrt {k^{2} + m_{e} ^{2}}}
\right)k^{2}dk} \quad ,\quad
n_{e}(\mu_{e})=\frac{\left(\mu_{e}^{2}-m_{e}^{2}\right)^{3}}{3\pi^{2}}.
\end{equation}

\noindent The condition of electrical neutrality for a $udse$
plasma is
\begin{equation}
\label{eq40} \frac{{2}}{{3}}n_{u} - \frac{{1}}{{3}}n_{d} -
\frac{{1}}{{3}}n_{s} - n_{e} = 0 \quad .
\end{equation}

Using the functions $n_{u} \left( {\mu ,\mu _{e}} \right)$, $n_{d}
\left( {\mu}  \right)$, $n_{s} \left( {\mu} \right)$, and $n_{e}
\left( {\mu _{e}} \right)$ from Eqs. (38) and (40), this equation
makes it possible to determine the function $\mu _{e} \left( {\mu}
\right)$ and, ultimately, the functions $\Omega _{u} \left( {\mu}
\right)\,,\;\,\Omega _{d} \left( {\mu} \right),\,\,\;\Omega _{s}
\left( {\mu} \right)$, and $\Omega _{e} \left( {\mu} \right)\,$.

The pressure of a "$udse$" plasma for a given chemical potential
$\mu$ is given by
\begin{equation}
\label{eq41} P_{QM} \left( {\mu}  \right) = - \sum\limits_{i =
u,{\kern 1pt} {\kern 1pt} d,{\kern 1pt} s,{\kern 1pt} e} {\Omega
_{i}}  \left( {\mu}  \right)\, - B\,,
\end{equation}

\noindent where $B$ is the "bag" constant, which characterizes the
vacuum pressure and ensures confinement. The energy density
$\varepsilon _{QM} $ and baryon concentration $n_{QM}$ of a
"$udse$" plasma are given by
\begin{equation}
\label{eq42} \varepsilon _{QM}(\mu) = \sum\limits_{i = u,{\kern
1pt} {\kern 1pt} d,{\kern 1pt} s,{\kern 1pt} e} {(\Omega _{i} +
\mu _{i} {\kern 1pt} n_{i}}) + \,B\,,
\end{equation}
\begin{equation}
\label{eq43} n_{QM}(\mu) = (n_{u} + n_{d} + n_{s})/3.
\end{equation}

Equations (41)-(43) give the equation of state of a quark-electron
("$udse$") plasma in the parametric form $\varepsilon _{QM}(P)$
and $n_{QM}(P)$. As in the case of an $npe$ -plasma, the baryon
chemical potential for quark-gluon matter is given by
\begin{equation}
\label{eq44} \mu _{QM}(P) = (P + \varepsilon _{QM}(P))/{n_{QM}(
P)}.
\end{equation}

\bigskip
\section{Phase transition to quark matter at constant pressure}
The modern concept of the phase transition between nuclear matter
and quark matter is based on a feature of this transition, first
noted by Glendenning \cite{Gl92,Gl00}, to the effect that there
are two conserved quantities in this transition: baryon number and
electrical charge. The requirement of global electrical neutrality
then leads to the possible formation of a mixed phase, where the
nuclear and quark matter are, separately, electrically charged,
while overall electrical neutrality is ensured by electrons
(leptons). In the case of a phase transition of this sort, the
energy density $\varepsilon $, baryon concentration $n$, and
chemical potential $\mu _{e}$ of the electrons, as well as the
pressure $P$, vary continuously. The question of whether the
formation of a mixed phase is energetically favorable, given the
finite dimensions of the quark structures inside nuclear matter,
the Coulomb interaction, and the surface energy, has been examined
elsewhere \cite{Hei93,Benh04,Vosk03,Mar07}. It was shown there
that the mixed phase is energetically favorable for small values
of the surface tension between the quark matter and the nuclear
matter. In this paper we assume that the transformation of nuclear
matter into quark matter is an ordinary first order phase
transition described by Maxwell's rule. A separate paper will deal
with the changes in the characteristics of the phase transition
with formation of a mixed phase \cite{Gl92} when the contribution
of a $\delta$ -meson field is included, as well as the influence
of these changes on the integral and structural parameters of
hybrid stars. In the case of an ordinary first order phase
transition, it is assumed that both nuclear and quark matter are
separately electrically neutral and that at some pressure $P_{0}$
corresponding to the coexistence of the two phases, the baryon
chemical potentials of the two phases are equal, i.e.,
\begin{equation}
\label{eq45} \mu _{NM} \left( {P_{0}}  \right) = \mu _{QM} \left(
{P_{0}}  \right).
\end{equation}

Note that the chemical potential per baryon in nuclear matter is
given by
\begin{equation}
\label{eq46} \mu _{NM}
=\left(\mu_{p}~n_{p}+\mu_{n}~n_{n}+\mu_{e}^{(NM)}~n_{e}^{(NM)}\right)/n\;,
\end{equation}
\noindent and in the case of neutral, $\beta$ -equilibrium nuclear
matter (because of the conditions $n_{p} - n_{e}^{\left( {NM}
\right)} = 0$ and $\mu _{p} = \mu _{n} - \mu _{e}^{\left( {NM}
\right)} $), coincides with the chemical potential $\mu _{n}$ for
a neutron given by Eq. (29). In the case of a neutral, $\beta$
-equilibrium quark-gluon plasma, the relationship between the
baryon chemical potential and the chemical potentials of a $d$
quark ($\mu _{d} = \mu$) and an electron ( $\mu _{e}^{\left( {QM}
\right)}$) has the form
\begin{equation}
\label{eq47} \mu _{QM} = 3\mu - \mu _{e}^{\left( {QM} \right)} ,
\end{equation}

\bigskip
\section{Numerical computations}
Table 3 lists the calculated phase transition parameters for the
"$\sigma \omega \rho \delta$+ MIT" model examined in this paper at
constant pressure (Maxwell rule) for 12 different values of the
"bag" parameter $B$.
\begin{table}
  \centering
  \caption{\small Parameters of a Maxwellian Phase Transition for Different
Values of the "Bag" Constant $B$ }\label{T3}
\begin{tabular}
[c]{|c|c|c|c|c|c|c|c|c|c|}\hline $B$ & $\mu_{b}$ & $n_{N}$ &
$n_{Q}$ & $P_{0}$ & $\varepsilon_{N}$ &
$\varepsilon_{Q}$ & $\mu_{e}^{(NM)}$ & $\mu_{e}^{(QM)}$ & $\lambda$\\
\small MeV/fm$^{3}$ & \small MeV & \small  fm$^{-3}$ & \small
fm$^{-3}$ &\small  MeV/fm$^{3}$ &\small  MeV/fm$^{3}$ &\small
MeV/fm$^{3}$ &\small  MeV &\small  MeV &
\\\hline
\small 60 & \small 965.9 & \small 0.1207 & \small 0.2831 & \small 2.11 & \small 114.5 & \small 271.4 & \small 99.14 & \small 9.205 & \small 2.327 \\
\small 65 & \small 999.7 & \small  0.1787 & \small 0.3161 & \small 7.22 & \small 171.4 & \small 308.8 & \small 138.0 & \small 8.350 & \small 1.728 \\
\small 69.3 & \small 1032 & \small 0.2241 & \small 0.3504 & \small 13.84 & \small 217.5 & \small 347.9 & \small 166.0 & \small 7.588 & \small 1.504\\
\small 70 & \small 1038 & \small 0.2312 & \small 0.3564 & \small 15.10 & \small 224.9 & \small 354.9 & \small 170.2 & \small 7.464 & \small 1.479\\
\small 75 & \small 1079 & \small 0.2810 & \small 0.4027 & \small 25.55 & \small 277.6 & \small 408.8 & \small 198.1 & \small 6.613 & \small 1.349\\
\small 80 & \small 1119 & \small 0.3276 & \small 0.4525 & \small 37.95 & \small 328.8 & \small 468.6 & \small 221.9 & \small 5.842 & \small 1.278\\
\small 85 & \small 1158 & \small 0.3704 & \small 0.5036 & \small 51.51 & \small 377.5 & \small 531.8 & \small 242.1 & \small 5.173 & \small 1.240\\
\small 90 & \small 1194 & \small 0.4089 & \small 0.5541 & \small 65.54 & \small 422.8 & \small 596.2 & \small 259.0 & \small 4.605 & \small 1.221\\
\small 95 & \small 1227 & \small 0.4435 & \small 0.6029 & \small 79.56 & \small 464.7 & \small 660.4 & \small 273.1 & \small 4.125 & \small 1.213\\
\small 100 & \small 1257 & \small 0.4746 & \small 0.6497 & \small 93.30 & \small 503.3 & \small 723.5 & \small 285.2 & \small 3.717 & \small 1.213\\
\small 110 & \small 1309 & \small 0.5281 & \small 0.7369 & \small 119.5 & \small 572.0 & \small 845.4 & \small 304.5 & \small 3.066 & \small 1.223\\
\small 120 & \small 1354 & \small 0.5729 & \small 0.8165 & \small 143.9 & \small 631.7 & \small 961.4 & \small 319.5 & \small 2.568 & \small 1.240\\
\hline
\end{tabular}
\end{table}
\begin{figure}[h]
\begin{center}
  \begin{minipage}[b]{0.48\linewidth}
   \begin{center}
   \includegraphics[height=0.9\linewidth, width=0.9\linewidth]%
    {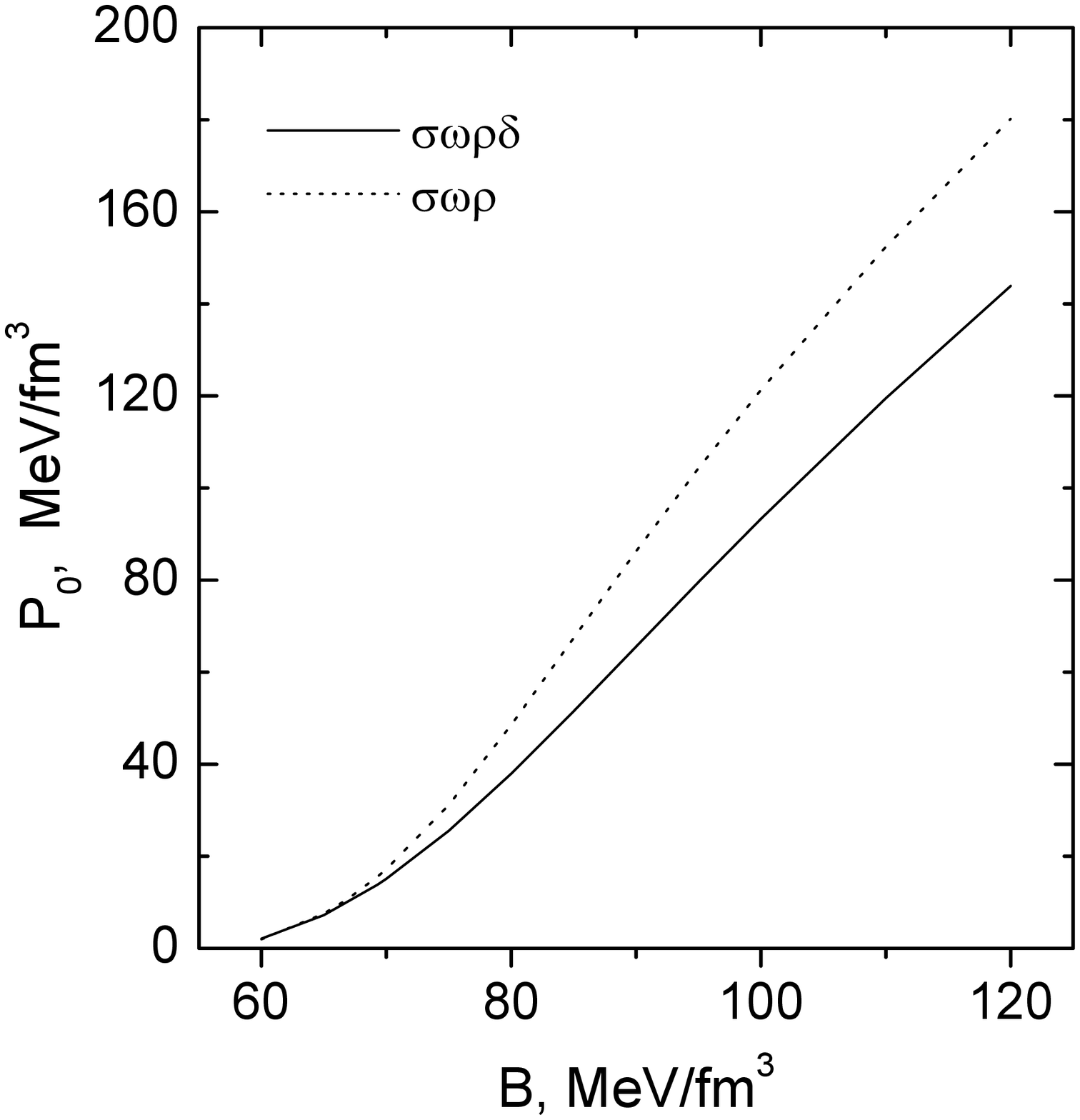}
    \caption {\small{ The phase transition temperature $P_{0}$ as a function of
the "bag" constant $B$. The smooth curve corresponds to the
"$\sigma \omega \rho \delta$" model and the dashed curve, to
"$\sigma \omega \rho$". }}
    \end{center}
  \end{minipage}\label{Fig6} \hfil\hfil
  \begin{minipage}[b]{0.48\linewidth}
   \begin{center}
   \includegraphics[height=0.9\linewidth, width=0.9\linewidth]%
    {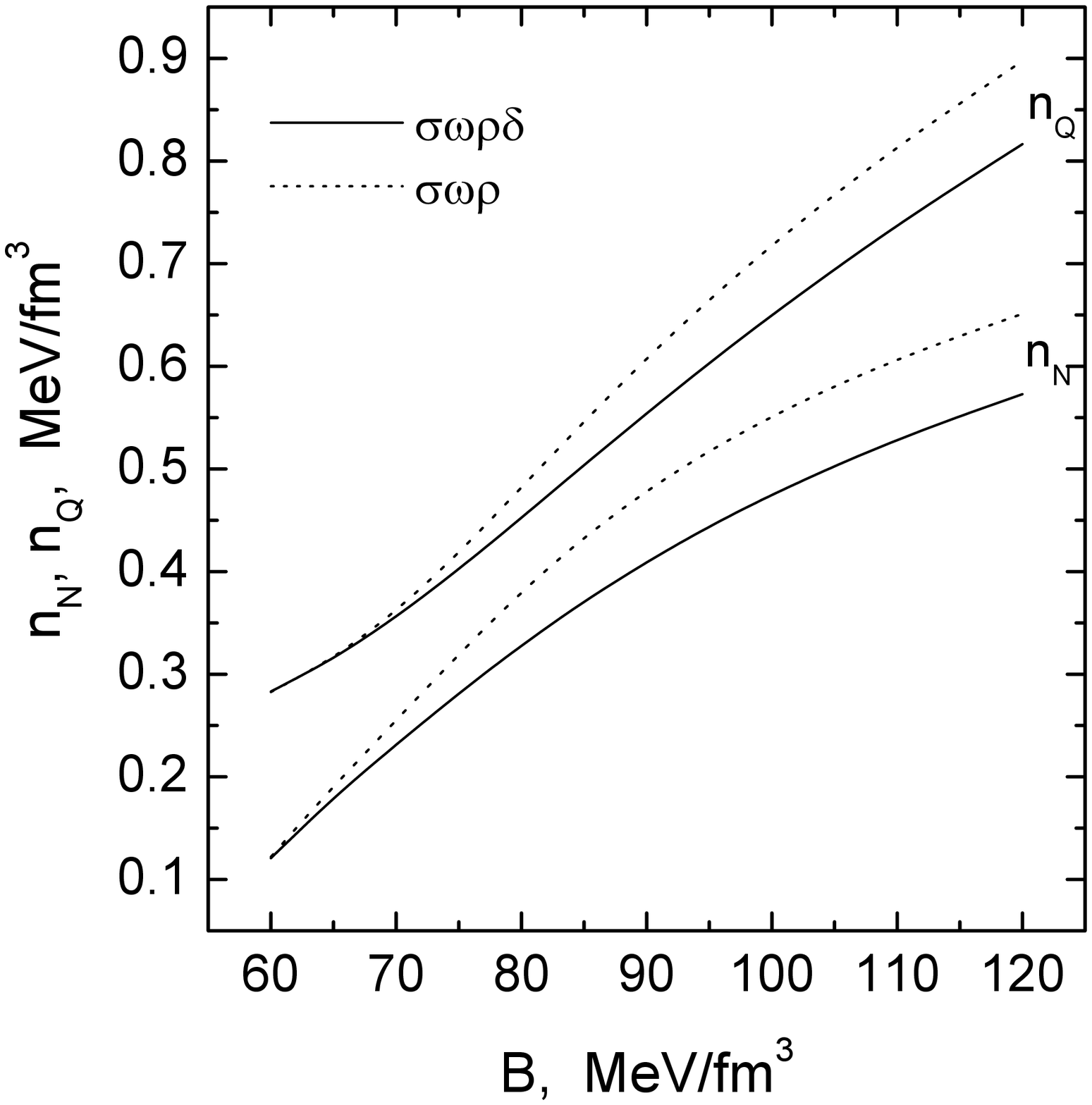}
    \caption {\small{Baryon concentrations of nuclear matter ($n_{N}$) and
strange quark matter ($n_{Q}$) at the point of a maxwellian phase
transition as functions of the "bag" constant $B$. Notation as in
Fig. 6.}}
    \end{center}
  \end{minipage}\label{Fig7}
 \end{center}
\end{figure}
\begin{center}
\begin{figure}[h]
 \begin{minipage}[t]{0.48\linewidth}
   \includegraphics[height=0.9\linewidth, width=0.9\linewidth]%
    {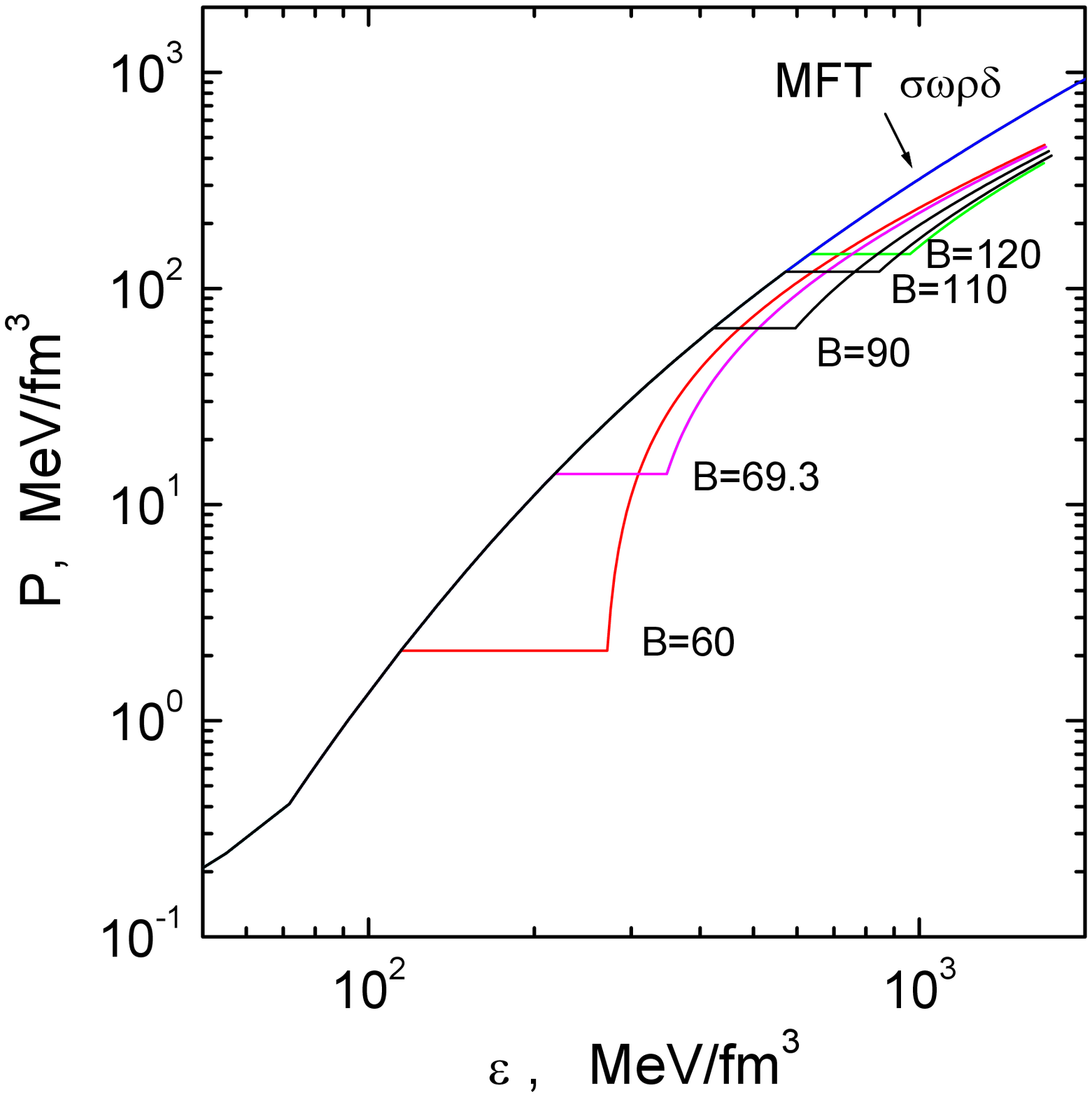}
    \caption {\small{ The equations of state for superdense matter with a
maxwellian phase transition calculated in the "$\sigma \omega \rho
\delta$" model for five different values of the parameter $B$. }}
  \end{minipage}\label{Fig8}
 \end{figure}
\end{center}
The quark masses are taken to be $m_{u} = 5$ MeV, $m_{d} = 7$ MeV,
and $m_{s} = 150$MeV, and the strong interaction constant, to be
$\alpha_{s}=0.5$. In this table $\mu _{b}$ is the baryon chemical
potential at the phase transition point, $n_{N}$ and $n_{Q}$ are
the baryon concentrations of the nuclear and quark matter,
respectively, at the transition point, $\varepsilon _{N} $ and
$\varepsilon _{Q}$ are the energy densities, $\mu _{e}^{\left(
{NM} \right)} $ and $\mu _{e}^{\left( {QM} \right)}$ the chemical
potentials of an electron in nuclear and quark matter,
respectively, and $P_{0}$ is the phase transition pressure.

It has been shown [28] that for a first order phase transition the
density discontinuity parameter
\begin{equation}
\label{eq48} \lambda = {{\varepsilon _{Q}}  \mathord{\left/
{\vphantom {{\varepsilon _{Q} } {\left( {\varepsilon _{N} + P_{0}}
\right)}}} \right. \kern-\nulldelimiterspace} {\left( {\varepsilon
_{N} + P_{0}}  \right)}}
\end{equation}
\noindent plays a decisive in the stability of neutron stars with
arbitrarily small cores made of matter from the second
(denser)phase. Paraphrasing the conclusions of that paper
\cite{Sei71}, in the case of a hadron-quark first order phase
transition, we have the following conditions. If
$\lambda\leq{3/2}$, then a neutron star with an arbitrarily small
core of strange quark matter is stable. On the other hand, for
$\lambda > 3/2$, neutron stars with small quark cores are
unstable. In the latter case, there is a nonzero minimum value of
the radius of the quark core for a stable star. Accretion of
matter to a neutron star when $\lambda > 3/2$ will lead to a
catastrophic (discontinuous) readjustment of the star, with
formation of a star that has a quark core of finite size. This
sort of catastrophic transition can also occur in the case of a
rotating neutron star that is slowing down, when the pressure in
the center rises and exceeds the threshold value, $P_{0}$. The
process of catastrophic readjustment with formation of a quark
core of finite radius at the star's center will be accompanied by
the release of a colossal amount of energy, comparable to the
energy release during a supernova explosion. The last column of
the table lists the discontinuity parameter $\lambda$ for the
various values of the "bag" constant $B$. The above mentioned
catastrophic readjustment of a neutron star (during accretion of
matter to its surface or as its rotation slows down) corresponds
to the first three versions of the equation of state listed in
Table 3, for which $B \le 69,3$ MeV/fm$^{3}$.

Figure 6 illustrates the dependence of the phase transition
pressure $P_{0}$ on the value of the "bag" parameter $B$. It is
clear that including a $\delta $-interaction channel leads to a
reduction in $P_{0}$. Similar plots of the baryon concentrations
of the nuclear ($n_{N}$) and quark ($n_{Q}$) phases at the phase
transition point are shown in Fig. 7. Evidently, including a
scalar-isovector effective $\delta$ -meson field reduces the
baryon concentrations for both phases at the phase transition
point. Then the density discontinuity parameter increases. Figure
8 shows the equation of state for superdense matter with a
maxwellian phase transition calculated in our "$\sigma \omega \rho
\delta$ + MIT" model for five different values of the parameter
$B$.

\bigskip

\section{Conclusion}

In this paper we have studied the equation of state of superdense
nuclear matter in terms of the relativistic mean-field theory,
including a scalar-isovector $\delta$ -meson effective field. The
values of the constants for the relativistic mean-field theory
that we have found have enabled us to calculate the
characteristics of asymmetric nuclear matter, as well as of
$\beta$ -equilibrium $npe$ -plasmas. The dependences of the
effective masses of protons and neutrons on the baryon
concentration n for given values of the asymmetry parameter   have
been studied and it has been shown that in an asymmetric nucleonic
medium the effective mass of a proton exceeds that of a neutron.

The dependence of the asymmetry parameter $\alpha$ on the baryon
concentration for $\beta$ -equilibrium $npe$ -plasmas has been
studied for different values of the electrical charge per baryon
and it was shown that including a $\delta$ -field reduces
$\alpha$.

Assuming that the phase transition between nuclear matter and
strange quark matter is an ordinary first order phase transition
obeying Maxwell's rule, we have made a detailed study of the
effect of including a $\delta$ -meson field on the parameters of
the phase transition. We have determined the phase transition
parameters for 12 different values of the bag parameter within the
range $B \in \left[ {60\,;\;120} \right]$ MeV/fm$^{3}$ and shown
that including a $\delta$ -meson field leads to reduction in the
phase transition pressure $P_{0}$ and in the concentrations
$n_{N}$ and $n_{Q}$ for coexistence of the two phases. Here the
density discontinuity parameter $\lambda$ increases. A bag
parameter $B \approx 69,3$ MeV/fm$^{3}$ corresponds to the
critical value $\lambda _{cr} = 3/2$. When $B < 69.3$ Mev/fm$^{3}$
the density discontinuity parameter obeys $\lambda > \lambda _{cr}
$ and neutron star configurations with infinitely small quark
cores will be unstable.

This analysis shows that a scalar-isovector $\delta$ -meson field
leads to more stiff equations of state for the nuclear matter
owing to splitting of the effective proton and neutron masses, as
well as to an increased asymmetry energy. It is known that a good
source of information on the rigidity of an equation of state for
dense matter is measurements of the mass of compact stars. The
mass of the compact star in a binary system associated with the
PSR pulsar $B1516+02B$ was recently measured and found to be $M =
2,08\; \pm 0,19M_{ \odot}$ \cite{Frei08}. The existence of neutron
stars with such high masses argues for a more stiff equation of
state than the equation which yields the standard value of $M =
1,44\;M_{ \odot}$.

Evidently, the above mentioned changes in the equation of state of
superdense matter and in the phase transition parameters will lead
to corresponding changes in both the structure and the integral
characteristics of hybrid stars with strange quark cores. A
separate article will be devoted to a study of the configuration
of neutron stars of this type, calculated by integrating the
system of Tolman-Oppenheimer-Volkoff equations based on the
equations of state obtained in this paper, with and without the
inclusion of a scalar-isovector effective $\delta$ -meson field.

\section*{Acknowledgements}
The author thanks Prof. Yu. L. Vartanyan for valuable advice and
support of the idea behind this article, as well as all the
participants in the scientific seminar at the department of the
theory of wave processes and physics in the Faculty of
Radiophysics at Yerevan State University for useful discussions.

This work was supported in the framework of topic 2008-130,
financed by the Ministry of Education and Sciences of the Republic
of Armenia.

\end{document}